\documentclass[prb,aps,twocolumn,noshowpacs,duperscriptaddresses]{revtex4-1}
\usepackage{graphicx}
\usepackage{dcolumn}
\usepackage{amsmath}
\usepackage{lmodern}
\usepackage{color}
\usepackage{longtable} 
\usepackage{units}
\usepackage{xcolor}

\definecolor{babyblue}{rgb}{0.54, 0.81, 0.94}

\newcommand{\tiem}[1]{{\text{\tiny{#1}}}}
\newcommand{\kB}{k_{\rm B}}
\newcommand{\TC}{T_{\rm C}} 

\newcommand{\mub}{\mu_{\mathrm B}}

\newcommand{\muiFE}{\mu_i^{\rm Fe}}                  
\newcommand{\muiNI}{\mu_i^{\rm Ni}}      
\newcommand{\muiCU}{\mu_i^{\rm Cu}}      
\newcommand{\muFE}{\mu^{\rm Fe}}                  
\newcommand{\muNI}{\mu^{\rm Ni}}      
      
\newcommand{\HFeNi}{{\cal H}}

\begin{document}

\widetext


\title{Multiscale modeling of ultrafast element-specific magnetization dynamics of ferromagnetic alloys 
}
\author{D.\ Hinzke$^1$}\email{denise.hinzke@uni-konstanz.de}
\author{U.\ Atxitia$^{1,2}$} 
\author{K.\ Carva$^{3,4}$} 
\author{P.\ Nieves$^5$}
\author{O. Chubykalo-Fesenko$^5$}
\author{P. M.  Oppeneer$^4$}
\author{U.\ Nowak$^1$} 
\affiliation{$^1$Fachbereich Physik, Universit\"{a}t Konstanz, D-78457 Konstanz, Germany}
\affiliation{$^2$Zukunftskolleg at Universit\"{a}t Konstanz, D-78457 Konstanz, Germany}
\affiliation{$^3$Faculty of Mathematics and Physics, DCMP, Charles University, Ke Karlovu 5, CZ-12116 Prague 2, Czech Republic}
\affiliation{$^4$Department of Physics and Astronomy, Uppsala University, Box 516, SE-751 20 Uppsala, Sweden}
\affiliation{$^5$Instituto de Ciencia de Materiales de Madrid, CSIC, Cantoblanco, 28049 Madrid, Spain}

\date{March 31, 2015}

\begin{abstract}
A hierarchical multiscale approach to model the magnetization dynamics of ferromagnetic random alloys is presented.  First-principles calculations of the Heisenberg exchange integrals are linked to atomistic spin models based upon the stochastic Landau-Lifshitz-Gilbert  (LLG) equation to calculate temperature-dependent parameters  (e.g., effective exchange interactions, damping parameters). These parameters are  subsequently used in the Landau-Lifshitz-Bloch (LLB) model for multi-sublattice magnets to calculate numerically and analytically the ultrafast demagnetization times. The developed multiscale method is applied here to FeNi  (permalloy) as well as to copper-doped FeNi alloys. \textcolor{black}{We  find that after an ultrafast heat pulse the Ni  sublattice demagnetizes faster than the Fe  sublattice for the here-studied FeNi-based alloys. }
\end{abstract}

\maketitle


\section{Introduction}

Excitation of magnetic materials by powerful femtosecond laser pulses leads to magnetization dynamics on the timescale of exchange interactions. 
For  elemental ferromagnets the emerging dynamics can be probed using conventional magneto-optical methods \cite{BeaurepairePRL1996ultrafast,KirilyukRMP2010ultrafast}. 
For magnets composed of several distinct elements, such as ferrimagnetic or ferromagnetic alloys, the individual spin dynamics of the different elements can be probed  employing ultrafast excitation in combination with the femtosecond-resolved x-ray magnetic circular dichroism (XMCD) technique  \cite{StohrBook2007magnetism,raduSPIE2012ultrafast}. 
An astonishing example of such element-specific ultrafast magnetization dynamics was first 
measured on ferrimagnetic GdFeCo alloys \cite{RaduNature2011transient}. There, it was observed that the rare-earth Gd sublattice demagnetizes in around 1.5 ps whereas  the transition metal FeCo sublattice has a much shorter demagnetization time of 300 fs.
Similar element-specific spin dynamics was also  observed  in CoGd and CoTb alloys\cite{LopezFloresPRB2013role,BergeardNatComm2014ultrafast}.
The element-selective technique allowed moreover to observe  for the first time the element-specific dynamics of the so-called ``all-optical switching"  (AOS) \cite{StanciuPRL2007AOS} in GdFeCo alloys,  finding that it unexpectedly proceeds through a transient-ferromagnetic-like state (TFLS) where  the FeCo  sublattice  magnetization points in the same direction
as that of the Gd  sublattice before complete reversal \cite{RaduNature2011transient,OstlerNatComm2012ultrafast}.
   Recent theoretical works  supported the  distinct demagnetization times observed experimentally \cite{AtxitiaPRB2014controlling,SchellekensPRB2013microscopic,WienholdtPRB2013orbital} and their crucial  role on the TFLS.
AOS has been also demonstrated for other rare-earth transition-metal ferrimagnetic alloys as TbFe 
\cite{HassdenteufelAdvMat2013thermally}, TbCo \cite{AlebrandAPL2012light}, TbFeCo \cite{ChengIEEETrans2012temperature}, DyCo \cite{ManginNatureMat2014engineered}, HoFeCo \cite{ManginNatureMat2014engineered},  synthetic ferrimagnets \cite{EvansAPL2014ultrafast,ManginNatureMat2014engineered,SchubertAPL2014all} and very recently in the hard-magnetic ferromagnet FePt \cite{LambertScience2014all}.  

Although the full theoretical explanation of the thermally driven AOS process is still a topic of debate \cite{OstlerNatComm2012ultrafast,MentinkPRL2012ultrafast,BarkerSREP2013two,WienholdtPRB2013orbital,BaryakhtarJETP2013exchange,AtxitiaPRB2013}, the distinct demagnetization rates of each of the constituting elements  has been suggested as the main driving mechanism for the AOS  observed on antiferromagnetically coupled  alloy \cite{OstlerNatComm2012ultrafast,AtxitiaPRB2014controlling,WienholdtPRB2013orbital}. These  findings have highlighted the question how ultrafast demagnetization would proceed in ferromagnetically coupled  two-sublattice materials  such as permalloy (Py). Unlike rare-earth transition-metal alloys which consists of two intrinsically different metals, Py is composed of Fe  (20 \%) and Ni (80 \%) which have a rather similar magnetic nature,
 due to a partially filled $3d$ shell.  Thus, it is \textit{a priori} not clear if their spin dynamics should be the same or different. 

Recent measurements   have addressed this question. Using extreme ultraviolet pulses from high-harmonic generation sources Mathias \textit{et al.}\cite{MathiasPNAS2012probing} probed  element-specifically the ultrafast demagnetization in Py and obtained the same demagnetization rates for each element, Fe and Ni, but with a 10 to 70 fs delay between them. 

From a theoretical viewpoint an important question is which materials parameter are defining for the ultrafast demagnetization.  Thus far,  different criteria have been suggested \cite{KazantsevaEPL2008slow,koopmansNatMater2010explaining}. For single-element ferromagnets, Kazantseva {\em et al.} \cite{KazantsevaEPL2008slow} estimated, based on phenomenological arguments, that the timescale for  the demagnetization processes is limited by $\tau_{\mbox{\small demag}} \approx \mu/(2\lambda\gamma \kB T_{\rm{pulse}})$. Here, $\tau_{\mbox{\small demag}}$  depends not only on the elemental atomic magnetic moment, $\mu$, 
but also on the electron temperature, $T_{\rm{pulse}}$, and on the damping constant $\lambda$. Assuming that the damping constants $\lambda$ and  gyromagnetic ratios $\gamma$ are equal for Fe and Ni the demagnetization time  would therefore only vary due to the different magnetic moments of the  constituting elements. In that case, the demagnetization time of Fe is larger than the one for Ni (since $\mu^{\rm Fe} > \mu^{\rm Ni}$, see  Table \ref{table:table1} below).

 A similar  criterion (as in Ref.\ \onlinecite{KazantsevaEPL2008slow} for single-element ferromagnets) has been suggested by Koopmans \emph{et al.} \cite{koopmansNatMater2010explaining} on the basis of the ratio between the magnetic moment and the Curie temperature, $\mu/\TC$.   
 Since for ferromagnetic alloys each element has the same Curie temperature, this  criterion would lead to the same conclusions as Kazantseva \textit{et al.}; the different atomic magnetic moments of Fe and Ni  are responsible for the different demagnetization times. 
Furthermore,  
 Atxitia \emph{et al.} \cite{AtxitiaPRB2014controlling} have theoretically estimated the demagnetization times in GdFeCo alloys proposing that the demagnetization times scale with the ratio of the magnetic moment to the exchange energy of each element and a similar relation is expected  for  ferromagnetic alloys.  The
 demagnetization times of Fe and Ni in Py were also theoretically investigated   by Schellekens and Koopmans in Ref.\
 \onlinecite{SchellekensPRB2013microscopic} where a modified microscopic three temperature model (M3TM)\cite{koopmansNatMater2010explaining} was used. Thereby, they obtained a perfect agreement with experimental results of Mathias \emph{et al.}, \cite{MathiasPNAS2012probing} but only assuming an at least 4 times larger damping constant for Fe. However, this work does not provide a simple general criterion, valid for other ferromagnetic alloys.

 We have developed a hierarchical multiscale approach  (cf.\ Ref.\ \onlinecite{KazantsevaPRB2008towards}) to investigate the
 element-specific spin dynamics of ferromagnetic alloys and to obtain a deeper insight into the underlying mechanisms. First, we construct and parametrize a model spin Hamiltonian for FeNi alloys on the basis of first-principles calculations [Sec.\ \ref{sec:atomisticA}]. This model spin Hamiltonian in combination with extensive  numerical atomistic spin  dynamics simulations based on the stochastic LLG equation are used to calculate the equilibrium properties  [Sec.\ \ref{sec:atomisticB}]  as well as the demagnetization process after the application of a step heat pulse. The second step of the presented multiscale model links the atomistic spin model to the macroscopic two-sublattices  Landau-Lifshitz-Bloch (LLB) equation of motion recently derived by Atxitia \emph{et al.} \cite{AtxitiaPRB2012Landau} [Sec.\ \ref{sec:LLB}]. The analytical LLB approach allows for efficient simulations, and most importantly, provides insight in the element-specific demagnetization times of FeNi alloys. 

\section{From first principles to atomistic spin model}
\label{sec:atomistic}
 \subsection{ Building the spin Hamiltonian}
 \label{sec:atomisticA}
 
 To start with, we construct an atomistic, classical spin Hamiltonian $\cal H$ on the basis of first-principles calculations. In particular, we consider three relevant alloys:
 Fe$_{50}$Ni$_{50}$, Fe$_{20}$Ni$_{80}$ (Py) and Py$_{60}$Cu$_{40}$. The first two alloys will allow us to  assess the influence of the Fe and Ni composition, while the last two alloys will permit us to study the effect of the inclusion of non-magnetic impurities on the demagnetization times.  
This was motivated by the work of Mathias \textit{et al.}\cite{MathiasPNAS2012probing} who studied the influence of Cu doping on the  Fe and Ni demagnetization times in an Py$_{60}$Cu$_{40}$ alloy.
  
 To obtain the spin Hamiltonian we have employed spin-density functional theory calculations to map the behavior of the magnetic material onto an effective Heisenberg Hamiltonian, which can be achieved in various ways \cite{liechtensteinJMMM1987local,Halilov98}.
Here we use
the two-step approach suggested by Lichtenstein \emph{et al.} 
\cite{LiechtensteinJPF1984exchange}.
The first step represents the calculation of the self-consistent electronic structure for a collinear spin structure at zero temperature. In the second step, exchange parameters of an effective classical Heisenberg Hamiltonian are determined using the one-electron Green functions. This method has been rather successful in explaining  magnetic thermodynamic properties of a broad class of magnetic materials 
\cite{TurekBOOK1997electronic,mryasovEL05,KudrnovskyPRB2008magnetic}.

The self-consistent electronic structure was calculated using the tight-binding linear muffin-tin orbital (TB-LMTO) approach \cite{TurekBOOK1997electronic} within the local spin-density approximation \cite{BarthSSP1972local} to the density functional theory.

Importantly, the materials we investigate here are alloys. Hence, it is assumed that atoms are distributed randomly on the host fcc lattice. The effect of disorder was described by the coherent-potential approximation (CPA) \cite{sovenPR1967coherent}. The same radii for constituent atoms were used in the TB-LMTO-CPA calculations. We have used around a million $k$-points in the full Brillouin zone  to resolve accurately energy dispersions close to the Fermi level.

\begin{figure}
\hspace*{-0.5cm}
\begin{centering}
\includegraphics[scale=0.7, angle = 0]{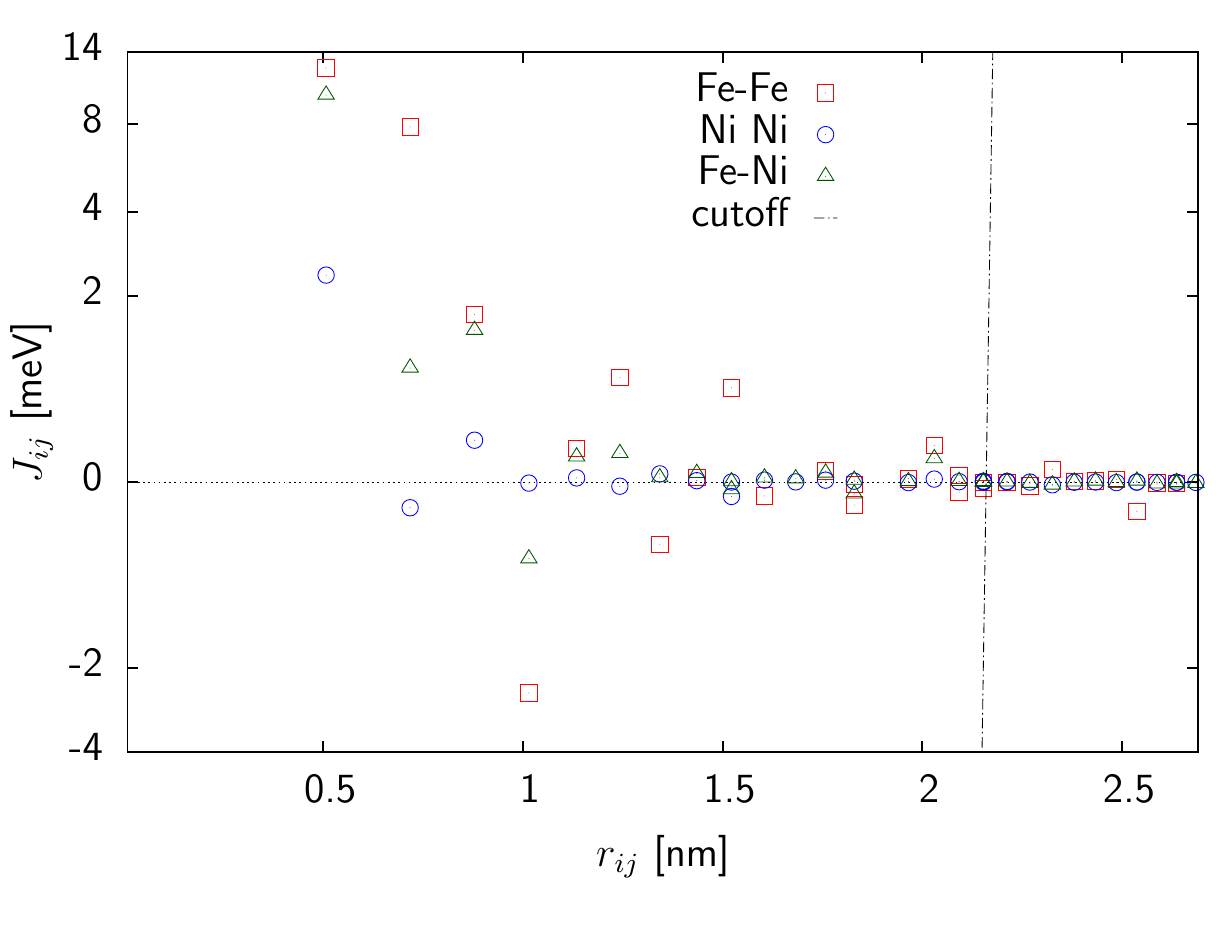}
\end{centering}
\caption{(Color online) \textit{Ab initio} calculated exchange constants $J_{ij}$ for the Fe$_{20}$Ni$_{80}$ alloy of the distance $r_{ij}$ between atoms $i$ and $j$. Results are given for the three different possible sublattice interactions ($J_{\rm Fe-Fe}$, $J_{\rm Ni-Ni}$, and $J_{\rm Fe-Ni}$). Note our hyperbolic scaling. In our atomistic spin simulations the exchange constants are taken into account up to a distance $r_{ij}$ (cutoff) where they are finally small enough to be neglected.}
\label{fig:exchange}
\end{figure}

The calculations of the Heisenberg exchange  constants $J_{ij}$ in ferromagnets can be performed with a reasonable numerical  effort by employing the magnetic force theorem \cite{LiechtensteinJPF1984exchange,liechtensteinJMMM1987local}. It allows to express the infinitesimal changes of the total energy using changes in one-particle eigenvalues due to non-self-consistent changes of the effective one-electron potential accompanying the infinitesimal rotations of spin quantization axes, i.e., without any additional self-consistent calculations besides that for the collinear ground state. The resulting pair exchange interactions are given by
\begin{equation}
J_{ij}=\frac{1}{\pi}\mathrm{Im} \!\! \int_{-\infty}^{E_{\rm{F}}} \! \! \! \!  \! \! \mathrm{d}E\int_{\Omega_{i}} \! \! \! \!  \mathrm{d}\mathbf{r}\int_{\Omega_{j}} \! \! \! \! \mathrm{d}\mathbf{r}'B_{\rm ex}\left(\mathbf{r}\right)G^{\uparrow}_{+} B_{\rm ex}\left(\mathbf{r'}\right)G^{\downarrow}_{-} ,
\end{equation}
 with $G^{\uparrow}_{+}=G^{\uparrow}\left(\mathbf{r},\mathbf{r'},E^{+}\right)$ and $G^{\downarrow}_{-} = G^{\downarrow}\left(\mathbf{r'},\mathbf{r},E^{-}\right)$.  $E_{\rm F}$ denotes the Fermi level and  $\Omega_{i}$ the $i$-th atomic cell,  $\sigma=\uparrow,\downarrow$
is the spin index, $E^{+}=\lim_{\alpha\rightarrow0}E+\mathrm{i}\alpha$, $G^{\sigma}$ are spin-dependent one-electron retarded Green functions, and $B_{\rm{ex}}$  is the magnetic field from the exchange-correlation potential.
The validity of this approximation has been examined more quantitatively in several studies. \cite{BrunoPRL2003exchange,katsnelsonJOP2004magnetic,turekPHILMAG2006exchange} The  \textit{ab initio} calculated distance-dependent exchange constants for the Fe$_{20}$Ni$_{80}$ alloy, i.e., the exchange within the Fe sublattice (Fe-Fe), the Ni sublattice (Ni-Ni) as well as between the Fe and Ni sublattices (Fe-Ni), are shown in Fig.\ \ref{fig:exchange}. The calculated magnetic moments  for all three alloys considered here are given in Table \ref{table:table1}.

 \begin{table*}
 \caption{ \textit{Ab initio} calculated magnetic moments $\mu^{\epsilon}$ and experimental lattice constants $\Delta$ used in the atomistic Langevin spin dynamics simulations. Effective exchange parameters calculated from  \textit{ab initio} calculations, $J^{\epsilon,\delta}_{0}=\sum_j J_{0j}^{\epsilon\delta}$, where the sum is here over all neighbors $j$. Curie temperatures as calculated from the atomistic simulations, $T_{\rm C}^{\rm{LLG}}$, and the experimental value, $T_{\rm C}^{\rm{exp}}$.
}
	\begin{center}
\begin{tabular}{ccccccccc}
\hline
\hline
 \\ [-2.5mm]
alloy &  $\muFE$ &  $\muNI$  & $\Delta$ & $J_0^{\mathrm{Ni-Ni}}$ & $J_0^{\mathrm{Fe-Fe}}$  &$J_0^{\mathrm{Fe-Ni}}$  &
$\TC^{\rm{LLG}}$ & $\TC^{\rm{exp}}$  \vspace{0mm} \\

&   $[\mub]$ &  $[\mub]$ & $\unit{[nm]}$ &   [J $ \times 10^{-21}$]&  [J$ \times 10^{-21}$] & [J$ \times 10^{-21}$] &
$[K]$ &  [K]    
 \\ [0.5mm] \hline
  \\ [-2.5mm]
Py &   2.637 &  0.628  & 0.3550 \cite{glaubitzJOP2011development} &
      $6.2419$ &  $32.3162$ & $26.3654$ & 
       650 &  850 \cite{MathiasPNAS2012probing}  \\
Ni$_{50}$Fe$_{50}$ & 
2.470 & 0.730 & 0.3588  \cite{abrikosovPRB2007competition} &
$6.6265$ & $25.3789$ 
  & $25.0656$ &  850  & \phantom{x}    \\
Py$_{60}$Cu$_{40}$ 
& 2.645 & 0.429 & 0.3550 \phantom{x} &
$2.6623$  & $56.2789$
 & $22.6442$  &  340 &  406 \cite{MathiasPNAS2012probing}  
 \\
\hline \hline
\end{tabular}
 \label{table:table1}
\end{center}
\end{table*}

In our hierarchical multiscale approach, these computed material parameters (the exchange constant matrix as well as the magnetic moments) are now used as material parameters for our numerical simulations based on an atomistic Heisenberg spin Hamiltonian.  We consider  thereto classical spins $\mathbf{S}_i^\epsilon = \boldsymbol{\mu}_i^{\epsilon}/\mu_i^{\epsilon}$   with $\epsilon$ randomly representing iron ($\mu^{\epsilon}_i$ = $\muiFE$) or nickel   magnetic moments ($\mu^{\epsilon}_i$ = $\muiNI$) on the fcc sublattice.  For the Cu-doped  Py$_{60}$Cu$_{40}$ alloy the calculated magnetic moments  on Cu vanish, {i.e.} $\muiCU = 0$.

The spin Hamiltonian for unit vectors, $\mathbf{S}_i^\epsilon$, representing the normalized magnetic moments of the $i$-th atom on either the Fe or Ni sublattice reads  
\begin{eqnarray}
  \label{eq:generic-Hamiltonian} 
        {\cal H} &=& -    \sum\limits_{ij} \Big( \frac{ J_{ij}}{2} \mathbf{S}_i^{\epsilon}  \cdot \mathbf{S}_j^{\delta} \\ \nonumber   &-& \frac{\mu_0 { \mu_i^{\epsilon} \mu_i^{\delta}}}{8 \pi}  \frac{{3 (\mathbf{S}_i^{\epsilon}\cdot {\mathbf e}_{ij})({\mathbf e}_{ij} \cdot \mathbf{S}_j^{\delta}) - \mathbf{S}_i^{\epsilon} \cdot \mathbf{S}_j^{\delta}}} {r_{ij}^3} \Big). 
                       \end{eqnarray}
The first sum represents  the exchange energy of magnetic moments, either on Ni or  on Fe sites, distributed randomly with  the required concentrations. The exchange interaction matrices  $J_{ij}$ (corresponding to $J_{\rm{Ni-Ni}}$, $J_{\rm{Fe-Ni}}$, or $J_{\rm{Ni-Ni}}$) are those from the \textit{ab initio} calculations (as shown for Py in Fig.\ \ref{fig:exchange}). These have been taken into account up to a distance of six unit cells (cutoff also shown in Fig.\ \ref{fig:exchange}) until they are finally small enough to be neglected.  The second sum describes the magnetic dipole-dipole coupling. 

Note, that the exchange interaction given by the matrices $J_{ij}$ is incorporated in our atomistic spin dynamics simulations via the Fast Fourier transformation method (see Ref.\ \onlinecite{HinzkeJMMM2000} for more details).
 As a side effect, we are able to calculate the dipolar interaction without any additional computational effort so that we take them into account although they will not  influence our results much.

Since we are interested in thermal properties we use Langevin dynamics, i.e.\ numerical solutions of the stochastic LLG equation of motion   
\begin{eqnarray}  
           \frac{ (1+(\lambda_i^{\epsilon})^2)\mu_i^{\epsilon}}{\gamma_i^{\epsilon}} {\dot{\bf{S}}}_i^{\epsilon} =
              -   \mathbf{S}_i^{\epsilon}  \times  \left[ \mathbf{H}_i 
              + \lambda_i^{\epsilon} \;   \left( \mathbf{S}_i^{\epsilon}  \times \mathbf{H}_i  \right) \right],
\end{eqnarray}
with the gyromagnetic ratio $\gamma_i^{\epsilon}$, and a dimensionless Gilbert damping constant $\lambda_i^{\epsilon}$  that describes the coupling to the heat-bath and corresponding either to Fe or to Ni. Thermal fluctuations are included as an additional noise term $\zeta_i$ in the internal fields $\mathbf{H}_i= - \frac{\partial \HFeNi}{\partial  \mathbf{S}_i^{\epsilon}} + \zeta_i(t)$ with  
\begin{equation}
\langle\zeta_i(t) \rangle = 0, \quad 
 \langle\zeta_{i\eta}(0) \zeta_{j\theta}(t) \rangle =\frac{2 \kB T \lambda_i^{\epsilon} \mu_i^{\epsilon}}{\gamma_i^{\epsilon}} \delta_{ij}\delta_{\eta\theta} \delta(t), 
\end{equation}
where  $i,j$ denotes lattice sites occupied either by Fe or Ni and  $\eta,\theta$  are Cartesian components.  All algorithms we use are described in detail in Ref.\ \onlinecite{NowakBOOK2007handbook}.


\subsection{Equilibrium properties: element-specific magnetization}
 \label{sec:atomisticB}
First, we investigate the  element-specific zero-field equilibrium magnetizations for Fe and Ni sublattices.  Those magnetizations are calculated as the spatial and time average of the sum of local magnetic moments, $\mathbf{m^\epsilon}=\langle \mathbf{S^\epsilon} \rangle$ with $\epsilon$ representing either Fe or Ni.  For our numerical studies, we assume  identical damping constants  ($\lambda = \lambda_i^{\epsilon}$) as well as  gyromagnetic ratios ($\gamma = \gamma_i^{\epsilon}$ = $1.76 \cdot 10^{11}$ (Ts)$^{-1}$) for both, Fe or Ni.  We perform our Langevin spin dynamics simulations for two different FeNi alloys, namely Fe$_{50}$Ni$_{50}$ and Py, as well as for permalloy diluted with copper, Py$_{60}$Cu$_{40}$.  All material parameters used in our simulations are given in Table \ref{table:table1}. 

\begin{figure}
\begin{centering}
\includegraphics[scale=1, angle = 0]{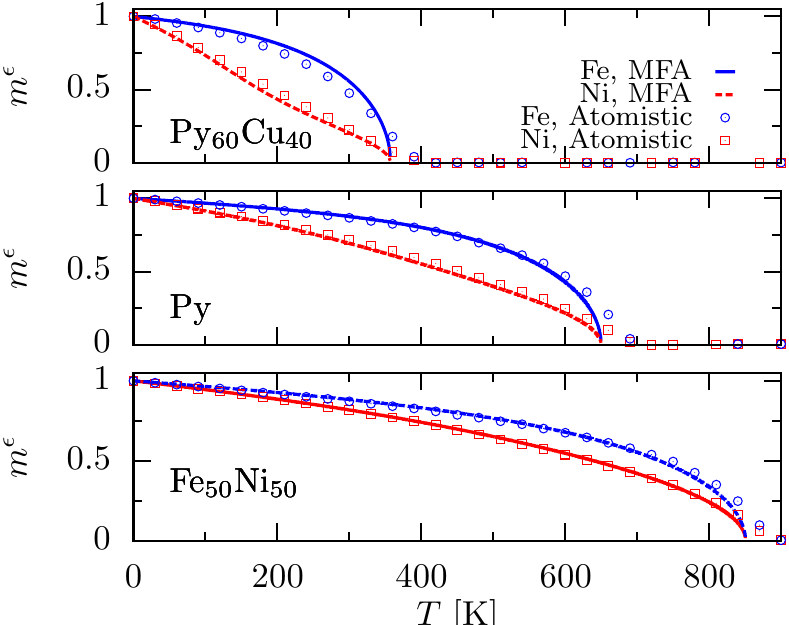}
\end{centering}
\caption{(Color online) Element-specific zero-field equilibrium magnetizations $m^{\epsilon}$ of either Fe or Ni as a function of temperature calculated by 
a rescaled mean-field approximation (MFA) (lines) and by the atomistic spin dynamics simulation (open symbols).  In the MFA the exchange parameters are renormalized by equalizing the Curie temperatures $\TC$ computed with atomistic simulations with those obtained from the rescaled MFA. System size 128 $\times$ 128 $\times$ 128, damping parameter $\lambda = 1.0$.}
\label{fig:M-T-Atomistic-MFA}
\end{figure}

 
 The temperature dependence of the normalized element-specific magnetizations $m^{\epsilon}$ are shown in Fig.\ \ref{fig:M-T-Atomistic-MFA}.  The calculated values of the Curie temperatures are given in Table \ref{table:table1} together with known experimental values. Both, the numerical and experimental values, are in good agreement. The element-specific magnetizations as well as the total magnetization (not shown in  Fig.\ \ref{fig:M-T-Atomistic-MFA}) of the alloys share the same Curie temperature
  while in the temperature range below the Curie temperature their temperature dependence is different for the two sublattices; the normalized magnetization of  Ni is lower than that of Fe.
  
The element-specific magnetizations calculated within the framework of a rescaled mean-field approximation (MFA) are shown as well. This approach will be discussed in detail in  Sec.\ \ref{sec:LLB}  below where these curves serve as material parameters for the simulations based on the LLB equation of motion also introduced in the next section. 

\section{From  atomistic spin model to macroscopic model}
\label{sec:LLB}

\subsection{Two-sublattices Landau-Lifshitz-Bloch equation}
Within the hierarchical multiscale approach, the  macroscopic (micromagnetic) equation of motion valid at elevated temperatures is the LLB equation \cite{KazantsevaPRB2008towards}. 
Initially, the macroscopic LLB equation of motion was derived by Garanin for single-species ferromagnets only. 
 Garanin  first  calculated
  the Fokker-Planck equation for a single spin coupled to a heat-bath, thereafter a non-equilibrium distribution function for the  thermal averaged spin polarization  was assumed to drive the non-equilibrium dynamics. 
 Second, the exchange interactions between atomic spins were introduced using the mean field approximation (MFA)
with respect to the spin-spin interactions.  This last step reduces to the replacement of the ferromagnetic spin Hamiltonian  $\cal{H}$ with the MFA Hamiltonian $\cal{H}_{\text{\tiny MFA}}$  in the single (macro)spin solution.
  
 The LLB formalism was recently broadened to describe the distinct dynamics
 of  two-sublattices magnets, both antiferromagnetically  or ferromagnetically coupled \cite{AtxitiaPRB2012Landau}. 
The derivation of such equations follows similar steps as for the ferromagnetic LLB version but considering sublattice specific  spin-spin exchange interactions  and  MFA exchange fields, $\left\langle \mathbf{H}^{\epsilon}_{\text{\tiny MFA}}\right\rangle^{\textrm{\tiem{conf}}}$. For the exchange field
  the random lattice model is used by generating the random average with respect to disorder configurations $\left\langle \dots \right\rangle^{\textrm{\tiem{conf}}}$.
The  corresponding set of coupled LLB equations for each sublattice  reduced magnetization
$\mathbf{m}^{\epsilon} = \langle \mathbf{S}^{\epsilon}\rangle =\mathbf{M}^{\epsilon}/M_{\rm{s}}^{\epsilon}$,  where $M_{\rm{s}}^{\epsilon}$ is the saturation magnetization at 0 K,
has  the form 
\begin{eqnarray}
\dot{\mathbf{m}}^{\epsilon} & =&  \gamma^{\epsilon}[\mathbf{m}^{\epsilon}\times\left\langle \mathbf{H}^{\epsilon}_{\text{\tiny MFA}}\right\rangle^{\textrm{\tiem{conf}}}]-
  \Gamma^{\epsilon}_{\bot}
  \frac{[\mathbf{m}^{\epsilon}\times[\mathbf{m}^{\epsilon}\times\mathbf{m}_{0}^{\epsilon}]]}{(m^{\epsilon})^{2}}
 \nonumber\\
  &-&\Gamma^{\epsilon}_{\|}\left(1-\frac{\mathbf{m}^{\epsilon}\mathbf{m}_{0}^{\epsilon}}{(m^{\epsilon})^{2}}\right)\mathbf{m}^{\epsilon}.
 \label{eq:LLBequationFeNi-NonEq}
 \end{eqnarray}
Here,  $\mathbf{m}_{0}^{\epsilon}=\mathcal{L}(\xi_{0}^{\epsilon})\frac{\boldsymbol{\xi}_{0}^{\epsilon}}{\xi_{0}^{\epsilon}}$  is  the transient (dynamical)  magnetization to which the non-equilibrium magnetization $\mathbf{m}^{\epsilon}$ tends to relax, and  where 
$\boldsymbol{\xi}_{0}^{\epsilon}\equiv \frac{\mu^{\epsilon}}{\kB T} \left\langle \mathbf{H}^{\epsilon}_{\text{\tiny MFA}}\right\rangle^{\textrm{\tiem{conf}}}$  is 
the thermal reduced field,
 $\xi_{0}^{\epsilon}\equiv\left|\boldsymbol{\xi}_{0}^{\epsilon}\right|$, and
 $\mathcal{L}\left(\xi\right)=\coth\left(\xi\right)-1/\xi$ is the Langevin function and $\mathcal{L}'(\xi) = \rm{d}\mathcal{L}(\xi)/\rm{d}\xi$. The parallel ($\Gamma^{\epsilon}_{\|}$) and perpendicular ($\Gamma^{\epsilon}_{\bot}$) relaxation rates in Eq. \eqref{eq:LLBequationFeNi-NonEq} are given by

\begin{equation}
 \Gamma^{\epsilon}_{\|}=\Lambda^{\epsilon}_{\rm N} \frac{1}{\xi_0^{\epsilon}}\frac{\mathcal{L}(\xi_{0}^{\epsilon})}{\mathcal{L}'(\xi_{0}^{\epsilon})}\quad \mbox{and} \quad \Gamma^{\epsilon}_{\bot}=\frac{\Lambda^{\epsilon}_{\rm N}}{2}\left(\frac{\xi_{0}^{\epsilon}}{\mathcal{L}(\xi_{0}^{\epsilon})}-1\right).
\label{eq:relaxations-rates-LLB}
\end{equation}
$\Lambda^{\epsilon}_{\rm N}=2 \kB T \gamma^{\epsilon}\lambda^{\epsilon}/ \mu^{\epsilon}$
is the characteristic diffusion relaxation
rate. 
The damping parameters $\lambda^{\epsilon}$ have the same origin as those 
 used in the atomistic simulations.

The first and the second terms on the right-hand side of Eq.\ \eqref{eq:LLBequationFeNi-NonEq} describe the 
transverse motion of the magnetization. These 
dynamics are much slower than 
the longitudinal magnetization dynamics given by the third term in this equation. Therefore, in the following we will neglect the
transverse components (in Eq.\ \eqref{eq:LLBequationFeNi-NonEq})
 and keep only the longitudinal one,
\begin{eqnarray}
\dot{m}^{\epsilon}  = -
\Gamma^{\epsilon}_{\|}  \left(m^{\epsilon}-m_{0}^{\epsilon} \right).
 \label{eq:LLBequationFeNi-NonEq-long}
 \end{eqnarray}
In spite of the fact that the form of Eq.\ \eqref{eq:LLBequationFeNi-NonEq-long} is similar to the well known Bloch equation, the quantity $m_0=m_{0}^{\epsilon}\left(m^{\epsilon},m^{\delta}\right)$  (with $\delta$ the 2-nd type of element) is not the equilibrium magnetization but changes dynamically through the dependence  of the effective field $\left\langle \mathbf{H}^{\epsilon}_{\text{\tiny MFA}}\right\rangle^{\textrm{\tiem{conf}}}$ on both sublattice magnetizations. Moreover, the rate parameter  $\Gamma^{\epsilon}_{\|}=\Gamma^{\epsilon}_{\|}\left(m_{0}^{\epsilon},m_{0}^{\delta}\right)$  contains highly non-linear terms in $m_{0}^{\epsilon}$ and $m_{0}^{\delta}$.

Therefore, the analytical solution of Eq.\ \eqref{eq:LLBequationFeNi-NonEq-long} and thus a deeper physical interpretation of the relaxation rates is difficult without any further approximations. However, Eq.\ \eqref{eq:LLBequationFeNi-NonEq-long} can be easily solved numerically with the aim to directly compare the solutions to those of the atomistic spin simulations. This is discussed in more detail in the next subsections. 

\subsection{From atomistic spin model to Landau-Lifshitz-Bloch equation}

\begin{figure}
\begin{centering}
\includegraphics[scale=0.4]{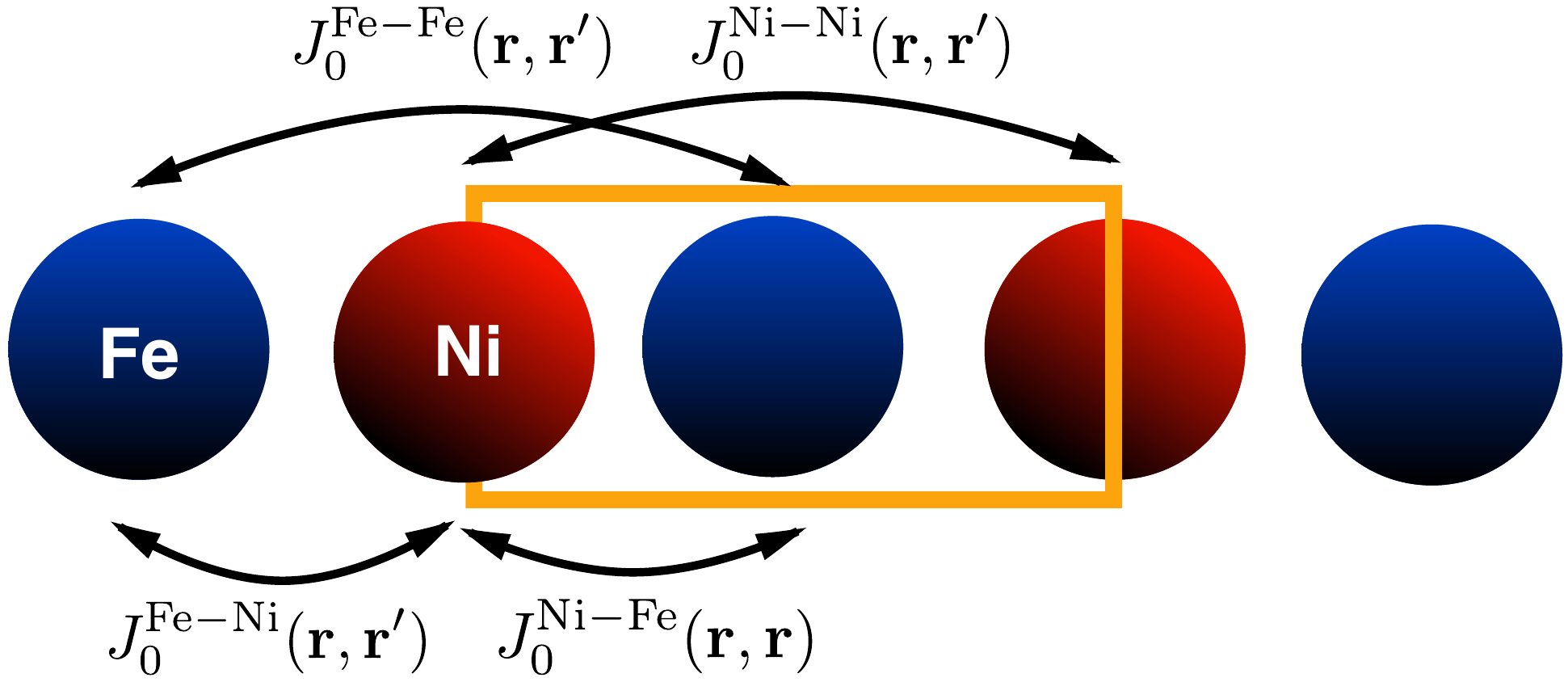}
\end{centering}
\caption{(Color online) Schematics of the magnetic unit cell used in the mean-field approximation for the FeNi alloys. The unit cell shown by the box contains two spins, one Fe and one Ni. The only  interaction among spins located at  the same unit cell $\mathbf{r}$ is defined by $J_{0}^{\rm{Ni-Fe}}$. The self-interactions are neglected, $J_{0}^{\rm{Ni-Ni}}(\mathbf{r},\mathbf{r})=J_{0}^{\rm{Fe-Fe}}(\mathbf{r},\mathbf{r})=0$. The rest of the interactions are among spins located in neighboring unit cells
$\mathbf{r}$ and $\mathbf{r}'$ .}
\label{fig:Schematics-FeNi-MFA-Jinteractions}
\end{figure}

Next, to solve Eq.\ \eqref{eq:LLBequationFeNi-NonEq} or Eq.\ \eqref{eq:LLBequationFeNi-NonEq-long}, one needs to calculate 
$\left\langle \mathbf{H}^{\epsilon}_{\text{\tiny MFA}}\right\rangle^{\textrm{\tiem{conf}}}$ for  the here-considered FeNi alloys. An adequate definition of such a field will allow us to directly compare  the magnetization dynamics from our atomistic spin simulation with the  LLB macroscopic approach. 

However,  a quantitative comparison between both a standard MFA and atomistic spin model calculations of the equilibrium properties is usually not  possible.  This is due to the fact that the Curie temperature gained with the MFA approach is overestimated due to the inherent poor approximation of the spin-spin correlations. 
Although,  rescaling the exchange parameters conveniently in such a way that the Curie temperature calculated with the MFA approach agrees with atomistic simulations leads to  a good agreement of both methods. Hence, we first present the standard MFA for disordered two-sublattices magnets, 
 thereafter, we will deal with the rescaling of the exchange parameters.

The MFA Hamiltonian of the full spin Hamiltonian  for FeNi alloys (see Eq.\ \eqref{eq:generic-Hamiltonian}
introduced in Sec.\ \ref{sec:atomistic}) can be written as 
%
\begin{equation}  
\mathcal{H}_{\mathrm{MFA}}=\mathcal{H}_{00}- \muFE
\sum_i \mathbf{H}_{\tiem{MFA}}^{\tiem{Fe}}\cdot \mathbf{S}^{\tiem{Fe}}_i-
\muNI \sum_i \mathbf{H}_{\tiem{MFA}}^{\tiem{Ni}}\cdot \mathbf{S}^{\tiem{Ni}}_i,
\label{eq:MFA-Hamiltonian-Py}
\end{equation}
where the dipolar interaction is neglected. The mean field acting on each site $i$ can be separated 
in two contributions; a) the contribution from neighbors of the same type $j^{\epsilon}$ and b) those of 
the other type $j^{\delta}$,
\begin{equation}
\mu^{\epsilon}
\left\langle \mathbf{H}^{\epsilon}_{\text{\tiny MFA}}\right\rangle^{\textrm{\tiem{conf}}}
=\sum_{\epsilon j^\epsilon} J^{\epsilon}_{j^{\epsilon}} \langle\mathbf{S}_{j^\epsilon} \rangle+\sum_{\epsilon j^\delta} J^{\epsilon}_{ j^\delta} \langle\mathbf{S}_{j^\delta} \rangle ,
\label{eq:MFA-Field-Py}
\end{equation}
where sums run over the nearest neighbours. When the homogenous magnetization approximation is applied (i.e.\
$\langle\mathbf{S}_{j^{\tiem{Fe}}} \rangle=\mathbf{m}^{\tiem{Fe}}$ and 
$\langle\mathbf{S}_{j^{\tiem{Ni}}} \rangle=\mathbf{m}^{\tiem{Ni}}$ for all sites) 
 one can define  $J_{0}^{\epsilon\epsilon}=\sum_{\epsilon j^{\epsilon}}J^{\epsilon}_{j^{\epsilon}}$ and $J_{0}^{\epsilon\delta}=\sum_{\epsilon j^{\delta}} J^{\epsilon}_{j^{\delta}}$. 
A sketch of the exchange interaction within the present MFA model is presented in Fig.\ \ref{fig:Schematics-FeNi-MFA-Jinteractions}. The impurity model is mapped to a regular spin lattice
where the unit cell (orange box) contains the two spin species, Fe and Ni, and the exchange interactions among them are weighted in terms of the concentration of each species.  
  
The equilibrium magnetization of each sublattice $m_{\rm e}^{\epsilon}$  can be obtained via the self-consistent solution of the Curie-Weiss equations 
$m_{\rm e}^{\epsilon}=\mathcal{L}(\frac{\mu^{\epsilon}}{\kB T} \left\langle \mathbf{H}^{\epsilon}_{\text{\tiny MFA}}\right\rangle^{\textrm{\tiem{conf}}})$. 

Fig. \ref{fig:M-T-Atomistic-MFA} shows
 good agreement of the calculated $m_{\rm e}^{\epsilon}(T)$ using the MFA and the atomistic spin model  for the three system studied in the present work. The exchange interactions  are rescaled as  $J_{0,\tiem{MFA}}^{\epsilon\delta}\simeq(1.65/2) J_{0}^{\epsilon\delta}$, for Fe$_{50}$Ni$_{50}$ and Py. 
For Py$_{60}$Cu$_{40}$ it is in agreement with 
 $J_{0,\tiem{MFA}}^{\epsilon\delta}=(1.78/2) J_{0}^{\epsilon\delta}$. Here, the atomistic calculations is not as accurate for intermediate temperatures as for the other two alloys. This could be because of the
increased complexity introduced by the inclusion of Cu impurities which cannot be fully described by the MFA.

\subsection{De- and remagnetization due to a heat pulse}
\begin{figure}
\hspace*{-2.5cm}
\begin{center}
\includegraphics[width = \columnwidth]{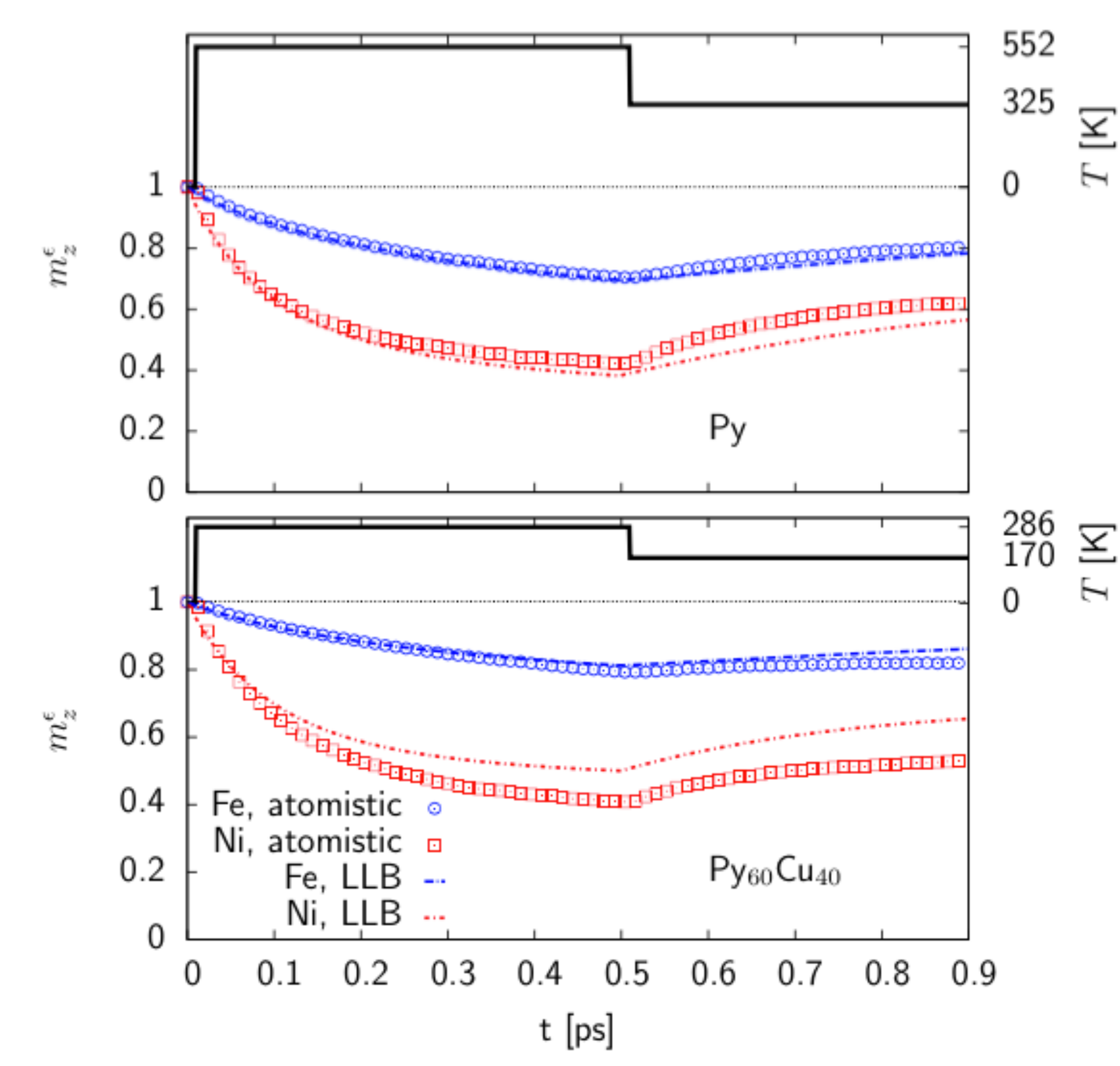}\\
\end{center}
\caption{(Color online) Calculated $z$-component of the normalized element-specific  magnetization $m_{z}^{\epsilon}$ vs.\ time for Py (top panel) and Py$_{60}$Cu$_{40}$ (bottom panel). In both cases the  quenching of the element-specific magnetizations for Fe and Ni due to a temperature step of  $T_{\rm{pulse}}$ = 0.8 $\TC$ are shown,  computed with atomistic Langevin spin dynamics (open symbols) as well as LLB simulations (lines). System size 64 $\times$ 64 $\times$ 64, damping parameter $\lambda = 0.02$.}
\label{fig:magPy}
\end{figure}

In the following, we study the reaction of the element-specific magnetization to a temperature step in Py as well as in Py diluted with Cu. In the first part of the temperature step the system is heated up to $  T = 0.8 \; \TC$ and in the second part it is cooled down to $T_{\rm{pulse}}$ =  0.5 \;$\TC$. The heat pulse roughly mimics the effect of heating due to a short laser pulse. The first part of the temperature step triggers the demagnetization while the second one triggers the remagnetization process.  We perform atomistic 
as well as LLB simulation of  the de- and remagnetization of the two sublattices after the application of a step heat pulse of 500-fs duration.

The reaction of the Fe and Ni sublattice magnetizations is shown in Fig.\ \ref{fig:magPy}. While the temperature step is switched on, the two sublattices relax to the corresponding equilibrium value of the sublattice magnetizations $m^{\epsilon}(T_{\rm{pulse}})$. Note, that these equilibrium values are different for the two sublattices in agreement with the temperature-dependent equilibrium element-specific magnetizations shown in Fig.\ \ref{fig:M-T-Atomistic-MFA}.

\begin{figure}
\begin{centering}
\includegraphics[scale=0.275, angle = 0]{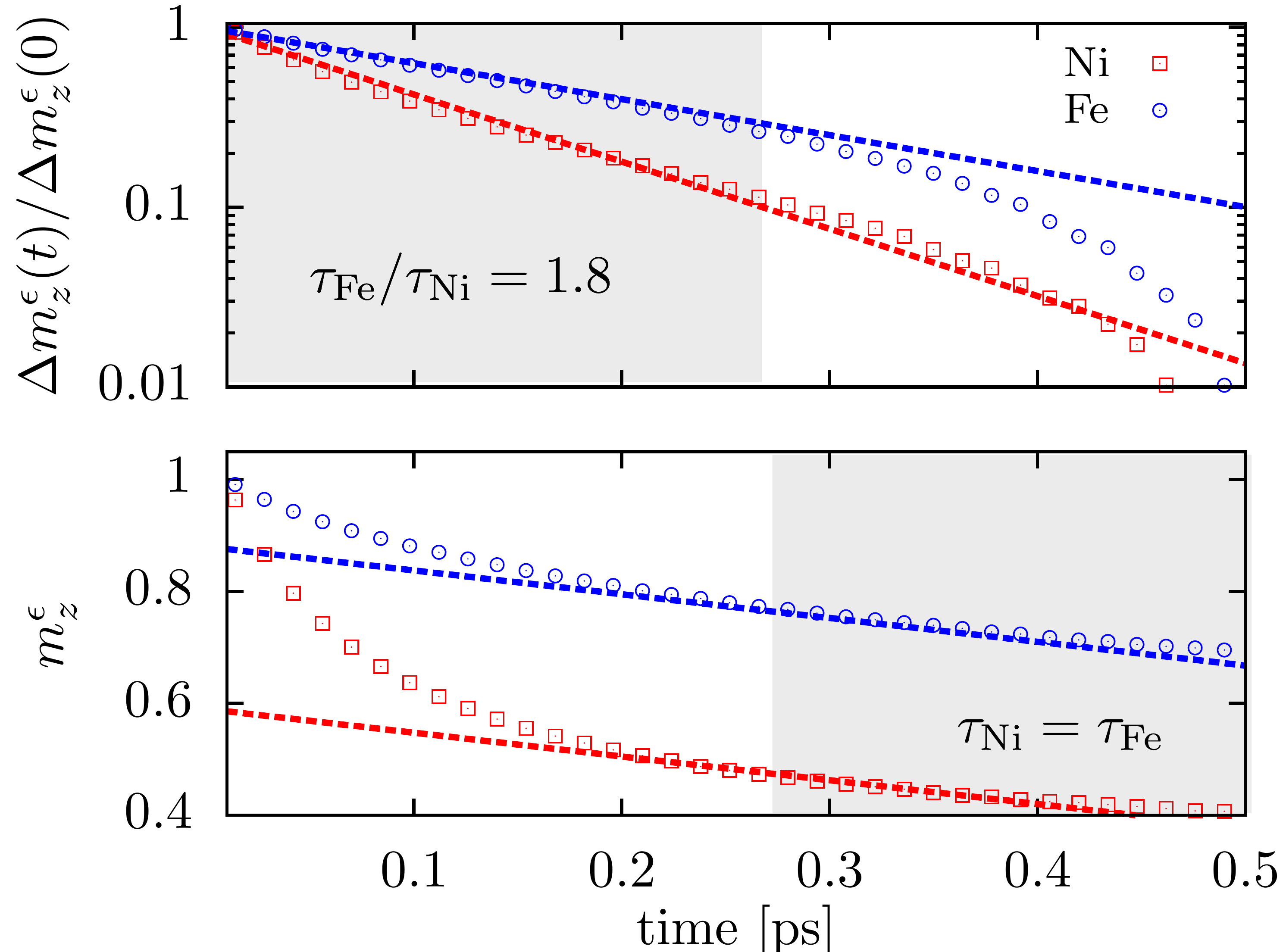}
\end{centering}
\caption{(Color online) Top panel: Normalized magnetization dynamics of Fe and Ni sublattices after the application of a heat pulse $T = 0.8 \;\TC$ as computed with the atomistic spin model.  The ratio between the Fe and Ni demagnetization times is 1.8. The intersection of the linear fit to the abscissa gives the relaxation time for each sublattice.  Bottom panel: plot of the unnormalized magnetization dynamics which shows that after the first 0.2 picosecond the element-specific demagnetization proceeds at the same rate.}
\label{fig:ExtractingRelaxTimeFromAtomistic}
\end{figure}

Because of that, the different demagnetization time scales are not well distinguishable in Fig.\ \ref{fig:magPy}. Thus, we use the normalized magnetization, 
$m_{\rm{norm}}^{\epsilon} = (m^{\epsilon}-m^{\epsilon}_{\rm{min}})/({m^{\epsilon}_{(t = 0)}-m^{\epsilon}_{\rm{min}}})$ of the sublattices, rather than $m^{\epsilon}$ to 
directly compare the demagnetization times.   
 \textcolor{black}{The demagnetization time after excitation with a temperature pulse is}\textcolor{black}{faster for Ni than for Fe} \textcolor{black}{(Fig.\ \ref{fig:ExtractingRelaxTimeFromAtomistic} (top panel)) for the first 200 fs, while one can see that for times larger than 200 fs both elements demagnetize at the same rate (Fig.\ \ref{fig:ExtractingRelaxTimeFromAtomistic} (bottom panel)).
Experiments on Py suggest that the time shift between distinct and similar demagnetization rates in Py is of around 10--70 femtoseconds \cite{MathiasPNAS2012probing}.}    

\subsection{Understanding relaxation times within the Landau-Lifshitz-Bloch formalism}
\label{sec:demag}

\begin{figure}
\begin{centering}
\includegraphics[scale=0.9, angle = 90]{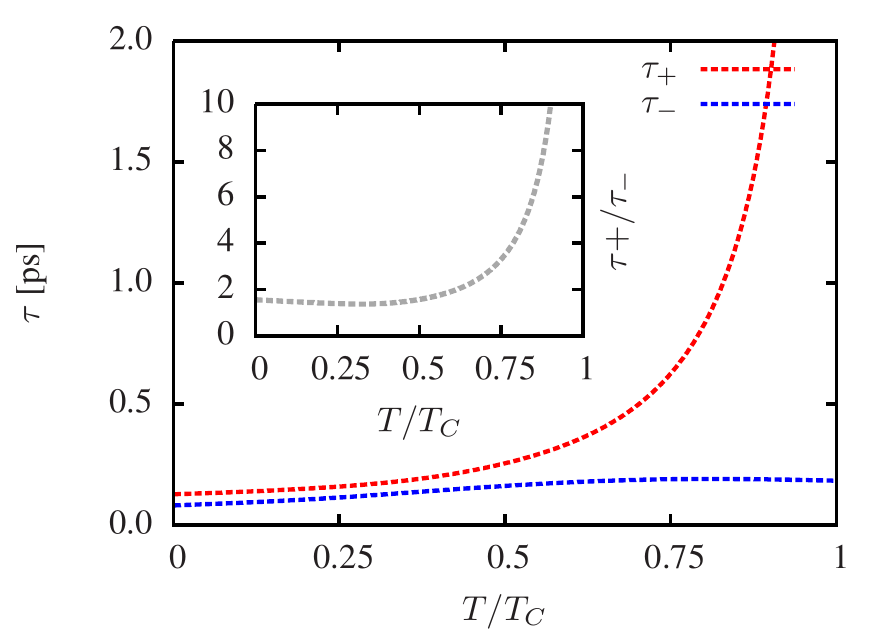}
\end{centering}
\caption{(Color online) Relaxation times of the dynamical system obtained by the LLB equation as a function of  temperature. Inset: The ratio between the relaxation times.}
\label{fig:FeNi-Longitudinal-Relaxation-Time}
\end{figure}

The relaxation rates of the Fe and Ni sublattices can be understood by discussing the linearized form of Eq.\ \eqref{eq:LLBequationFeNi-NonEq-long}. Here, the expansion of $\Gamma^{\epsilon}_{\|}$ and $m_{0}^{\epsilon}$ around their equilibrium values $m_e^{\epsilon}$  is considered  \cite{AtxitiaPRB2012Landau}  and leads to  $\partial (\Delta \mathbf{m}) / \partial t =\mathcal{A}_{\|}\Delta \mathbf{m}$ with  $\Delta\mathbf{m}=(\Delta m^{\epsilon}, \Delta m^{\delta})$ and $m^{\epsilon(\delta)}=m_{e}^{\epsilon(\delta)}+\Delta m^{\epsilon(\delta)}$. Furthermore, the  characteristic matrix $\mathcal{A}_{\|}$ drives the dynamics of this linearized equation and has the form 
\begin{equation}
\label{matrix:matrixFeNiLongitudinal}
\mathcal{A}_{\|}=\left(\begin{array}{ccc}
 -\gamma^{\epsilon}\alpha_{\|}^{\epsilon}/\Lambda^{\epsilon\epsilon}
 \enskip   &
 \gamma^{\epsilon} \alpha_{\|}^{\epsilon}
  J_{0}^{\epsilon\delta}/\mu^{\epsilon}
   \\
 \gamma^{\delta} \alpha_{\|}^{\delta}
 J_{0}^{\delta\epsilon}/\mu^{\delta} \enskip
  & -\gamma^{\delta}\alpha_{\|}^{\delta}/\Lambda^{\delta\delta}
 \end{array}\right) ,
\end{equation}
with
\begin{equation}
\Lambda^{\epsilon\delta} = \frac{J_{0}^{\epsilon\delta}}{\mu^{\epsilon}} \frac{m_e^{\epsilon}}{m_{\rm e}^{\delta}} \quad \quad \textrm{and} \quad \quad
\Lambda^{\epsilon\epsilon}=\frac{\widetilde{\chi}^{\epsilon}_{\|}} 
{1+\frac{J_{0}^{\epsilon\delta}}{\mu^{\epsilon}}
\widetilde{\chi}^{\delta}_{\|}} ,
\label{eq:Effective-suscep}
\end{equation}
where $\widetilde{\chi}^{\epsilon}_{\|}$ are the longitudinal susceptibilities which can be evaluated in the MFA approximation as
%
%

\begin{equation} 
\widetilde{\chi}^{\epsilon}_{\|}=   \frac{J_{0}^{\epsilon\delta} \mu^{\delta}\mathcal{L}^\delta \mathcal{L}^{\epsilon}
+\mu^{\epsilon}\mathcal{L}^{\epsilon}(\kB T- J_{0}^{\delta} \mathcal{L}^{\delta})}{ (\kB T- J_{0}^{\delta} \mathcal{L}^{\delta})(\kB T-J_{0}^{\epsilon} \mathcal{L}^{\epsilon})-J_{0}^{\epsilon\delta}J_{0}^{\delta\epsilon} \mathcal{L}^{\delta}\mathcal{L}^{\epsilon}},
\label{eq:susceptibilities}
\end{equation}
with $\mathcal{L}^{\epsilon} = \mathcal{L}^{'}(\xi_{\rm e}^{\epsilon})$ and  $\mathcal{L}^{\delta} = \mathcal{L}^{'}(\xi_{\rm e}^{\delta})$.
 We note that the longitudinal susceptibility 
 in Eq.\ \eqref{eq:susceptibilities} depends on the exchange parameter (Curie temperature)
and the atomic magnetic moments of both sublattices.

Next, the longitudinal damping parameter in Eq.\ \eqref{matrix:matrixFeNiLongitudinal} is defined as $\alpha^{\epsilon}=(2 \kB T\lambda^{\epsilon}m_e^{\epsilon})/\mu^{\epsilon}H_{e,{\rm ex}}^{\epsilon}$, where  $H_{e,{\rm ex}}^{\epsilon}$ is the average exchange field for the sublattice $\epsilon$ at equilibrium, defined by the MFA expression \eqref{eq:MFA-Field-Py}. The longitudinal fluctuations  are defined by the exchange energy, according to the expression above.
However, the longitudinal relaxation time is not simply inversely proportional to the damping parameter. Instead the relaxation parameters in Eq.\ \eqref{matrix:matrixFeNiLongitudinal} do also depend on the longitudinal susceptibilities which give the main contribution to their temperature dependence.

It is important to note that the matrix elements in Eq. \eqref{matrix:matrixFeNiLongitudinal} are temperature as well as (sublattice) material parameter dependent.  The general
solution of the characteristic equation, $|\mathcal{A}_{\|}-\Gamma^{\pm}
\mathcal{I}| =0$, gives two different eigenvalues, $\Gamma^{\pm}=1/\tau_{\pm}$,
corresponding to the eigenvectors $\mathbf{v}_{\pm}$. Here, $\mathcal{I}$ is the unit matrix. The computed temperature dependence of the relaxation times $\tau_{\pm}$ is presented in Fig.\ \ref{fig:FeNi-Longitudinal-Relaxation-Time}. 
 \textcolor{black}{More interestingly, we observe that the ratio between relaxation times $\tau_+/\tau_-$ [inset Fig. \ref{fig:FeNi-Longitudinal-Relaxation-Time}] is almost constant for temperature below $0.5 \;\TC$ and it has a value of 1.8 which compares well with atomistic simulations [Fig. \ref{fig:ExtractingRelaxTimeFromAtomistic}]. At elevated temperatures, one relaxation time $\tau_+$  will dominate the magnetization dynamics of both sublattices.}

\begin{figure}
\begin{center}
\includegraphics[scale=0.7]{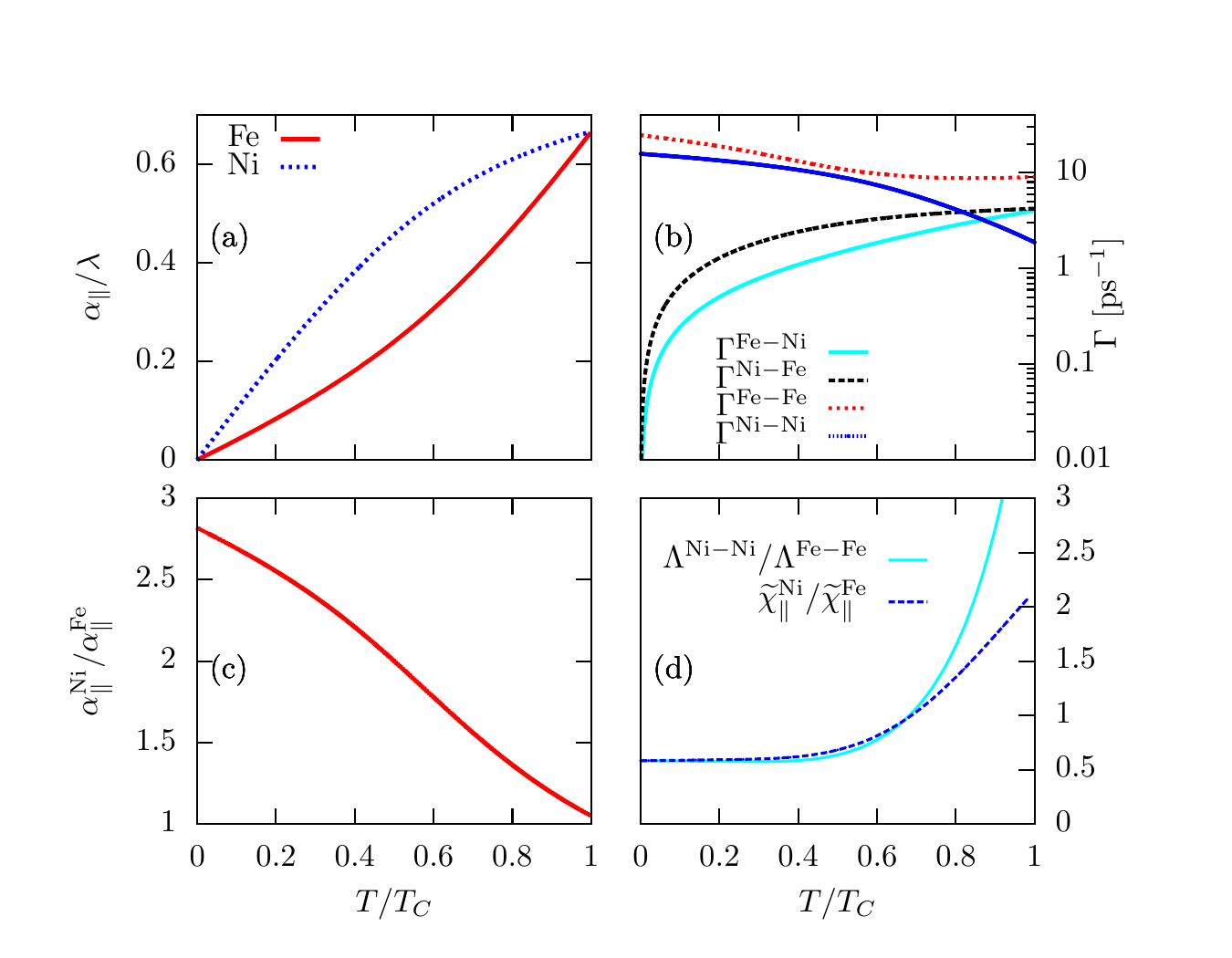}
\end{center}
\caption{(Color online) (a) Temperature dependence of the individual longitudinal damping parameters for Fe and Ni. (b) Matrix elements of the dynamical system defining   the magnetization dynamics. (c) Ratio between the individual damping parameters. (d) Ratio between the ``effective" susceptibilities $\Lambda^{\epsilon\epsilon}$ and the actual susceptibilities $\widetilde{\chi}_{\|}^{\epsilon}$.}
\label{fig:LLB-FeNi-parameters}
\end{figure}

  In Fig.\ \ref{fig:LLB-FeNi-parameters}(a)  we present the temperature dependence of the longitudinal damping parameters and in Fig.\ \ref{fig:LLB-FeNi-parameters}(b) the temperature dependence of the parameters $\Gamma^{\epsilon\delta}=\alpha_{\|}^{\epsilon}/\Lambda^{\epsilon\delta}$. These parameters  
 define the element-specific longitudinal dynamics.  In Figs.\ \ref{fig:LLB-FeNi-parameters} (c)  and (d)  the temperature dependent $\alpha_{\|}^{\epsilon}/\alpha_{\|}^{\delta}$ and  $\Lambda^{\epsilon\epsilon}/\Lambda^{\delta\delta}$ are shown. It can be seen that at least in the range of low temperatures the magnetization dynamics is mainly defined by $\Gamma^{\epsilon\epsilon}\gg \Gamma^{\epsilon\delta}$.

The general solution of the linearized LLB system for the two sublattices
can be written as 
%
\begin{eqnarray}
\Delta m^{\textrm{Fe}}(t) &=&
A^{\textrm{Fe}} \exp{\left(- t/\tau_+\right)}
+B^{\textrm{Fe}}\exp{\left(-t/\tau_-\right)} \nonumber \\
\Delta m^{\textrm{Ni}}(t) &=&
A^{\textrm{Ni}} \exp{\left(- t/\tau_+\right)}
+B^{\textrm{Ni}}\exp{\left(-t/\tau_-\right)}, 
\end{eqnarray}
where the coefficients $A^{\textrm{Fe(Ni)}}$
and $B^{\textrm{Fe(Ni)}}$ will depend of the eigenvectors $\mathbf{v}_\pm$ and the initial magnetic state, $\Delta m^{\textrm{Fe}}(0)$ and $\Delta m^{\textrm{Ni}}(0)$. For instance 
\begin{equation}
A^{\textrm{Fe}}=
\Delta m^{\textrm{Fe}}(0) 
\frac{\left[
1 -\frac{\Delta m^{\textrm{Ni}}(0)}{\Delta m^{\textrm{Fe}}(0)}  x_+
\right]
x_-}{x_--x_+}  ,
\label{eq:CoefficientA}
\end{equation}
where $x_+=v_+^{\textrm{Fe}}/v_+^{\textrm{Ni}}$ and $x_-=v_-^{\textrm{Fe}}/v_-^{\textrm{Ni}}$, is the ratio between he eigenvector components. 
The other coefficients are calculated similarly.  This complexity prohibits a general  analysis of the results. Thus, 
although the general solution is clearly a bi-exponential decay, one can wonder when the one exponential decay  approximation will give a good estimate 
for the individual relaxation dynamics.

Two interesting scenarios exist: First, the relaxation times 
$\tau_{+}$ and $\tau_-$  could have very different time scales and thus  
 one can separate the solution on short and long time scales, defined by 
$\tau_-$ and $\tau_+$, respectively. This is an interesting scenario for 
ultrafast magnetization dynamics where only the fast time scale will be relevant. 
Fig.\ \ref{fig:FeNi-Longitudinal-Relaxation-Time} shows the ratio $\tau_+/\tau_-$ and we can observe that the scenario $\tau_+/\tau_ -\gg 1$ only happens for temperatures approaching $\TC$. 
\textcolor{black}{As we have seen in the atomistic simulations, 
after an initial distinct quenching of each sublattice magnetization, both sublattice demagnetize at the same rate but slower than the initial rates (see Fig.\ \ref{fig:ExtractingRelaxTimeFromAtomistic}). }

The second scenario occurs when   
 $A^{\textrm{Fe}}\approx \Delta m^{\textrm{Fe}}(0)$ 
and $B^{\textrm{Ni}}\approx \Delta m^{\textrm{Ni}}(0)$, 
 even if $\tau_+$ and $\tau_-$ are 
of the same order. 
This happens, for example, either when the coupling between sublattices  is very weak, or at relatively low temperatures, see  Fig.\ \ref{fig:FeNi-Longitudinal-Relaxation-Time}. In this case 
 the system can be considered as two uncoupled ferromagnets (although with renormalized parameters), meaning that the matrix in Eq.\ \eqref{matrix:matrixFeNiLongitudinal} defining the dynamics is almost diagonal. 
Thus, we can approximately associate each eigenvalue of Eq.\ \eqref{matrix:matrixFeNiLongitudinal} 
to each sublattice, $\tau_{-}=\tau^{\textrm{Ni}}$ and 
$\tau_{+}=\tau^{\textrm{Fe}}$.  The inset in Fig.\ \ref{fig:FeNi-Longitudinal-Relaxation-Time} shows the ratio $\tau_+/\tau_{-}$ for the whole range of temperatures. At low-to-intermediate temperatures we find that  
$\tau_+/\tau_{-} \approx 1.8$.  \textcolor{black}{This is in good agreement with atomistic simulations, see Fig.\ \ref{fig:ExtractingRelaxTimeFromAtomistic}(a), and it clearly shows that the relaxation times ratio is not defined by the ratio between atomic magnetic moments,
$\muFE/\muNI\approx 4$.}

 In the case that the longitudinal relaxation rates are defined by the diagonal elements of the matrix \eqref{matrix:matrixFeNiLongitudinal} and $T$ is not close to $\TC$ the longitudinal relaxation time can be estimated as
 \begin{equation}
 \tau^\epsilon\simeq\frac{1}{2\gamma^{\epsilon}\lambda^{\epsilon}m_e^{\epsilon}H_{e,{\rm ex}}^{\epsilon}} .
 \label{relax}
 \end{equation}
Thus the ratio between the relaxation rates of Ni and Fe  (for the same gyromagnetic ratio value, the same coupling parameter and not too close to $\TC$)  is defined by
\begin{equation}
\frac{\tau^{\rm{Ni}}}{\tau^{\rm{Fe}}}=\left(
\frac{\lambda^{\rm{Fe}}}{\lambda^{\rm Ni}} 
\frac{\muNI}{\muFE} 
\right)
\frac{\widetilde{J}_0^{\rm Fe}m_e^{\rm Fe}}
{\widetilde{J}_0^{\rm Ni}m_e^{\rm Ni}}.
\label{eq:ratio-demag-time-FeNi}
\end{equation}
We recall that  $\widetilde{J}_{0}^{\epsilon}m_e^{\epsilon}=J_{0}^{\epsilon}m_e^{\epsilon}+J_{0}^{\epsilon\delta}m_e^{\delta}$ is the average exchange energy for the sublattice $\epsilon$ at equilibrium. Thus, the interpretation of the ratio of the relaxation times is straightforward. 
 The low temperature value of  
the ratio $\widetilde{J}_{0}^{\textrm{Fe}}/\widetilde{J}_{0}^{\textrm{Ni}}$ 
 is presented  in Table  \ref{table:table2} for the three alloys studied here. The second column  presents  the ratio between atomic magnetic moments, and the third column the estimated ratio between relaxation times under the assumption of equal damping parameter at each sublattice.  

\begin{table}[t!]
\caption{  Theoretical results: \emph{ab initio} calculated ratio between the mean exchange interaction at $T=0$ K,  the ratio between atomic magnetic moments and the quotient of these ratios. Results of simulations:  atomistic spin model calculated ratio between $\kappa$ exponents and relaxation times.  The ratio between the magnetic atomic moments and the exponents $\kappa$ is predicted in the main text to give the ratio between relaxation times.}
 \begin{ruledtabular}
\begin{tabular}{ lccccccc}
   \vspace{1.mm}
   &  \multicolumn{3}{c}{theoretical}   &  \multicolumn{3}{c}{simulations}  \\
      \cline{2-4} \cline{5-7}\\
   \vspace{1.5mm}
  alloy &   
 $\frac{\widetilde{J}_0^{\textrm{Fe}}}{\widetilde{J}_0^{\textrm{Ni}}}$ & $\frac{\muFE}{\muNI}$ & $\frac{\muFE}{\muNI} \frac{\widetilde{J}_0^{\textrm{Ni}}}{\widetilde{J}_0^{\textrm{Fe}}}
$
& 
$\frac{\kappa^{\textrm{Fe}}}{\kappa^{\textrm{Ni}}}$ & $\frac{\tau^{\textrm{Fe}}}{\tau^{\textrm{Ni}}}$ & $\frac{\muFE}{\muNI}
 \frac{\kappa^{\textrm{Fe}}}{\kappa^{\textrm{Ni}}}
$
  \\ \hline \hline
Fe$_{50}$Ni$_{50}$ & 1.592 & 3.38 & 2.12 & 
1.492 & 2.10 & 2.25
\\ 
Py                             & 2.685 & 4.198 & 1.563 &
2.3 & 1.8 & 1.8  \\ 
Py$_{60}$Cu$_{40}$  & 4.412 & 6.17 & 1.398
& 2.95 & 2.1 & 2.05 
 \\ 
 \end{tabular}
 \end{ruledtabular}
    \vspace{1.5mm}
 \label{table:table2}
 \end{table}
The estimated ratios for  relaxation times are in rather good agreement with the atomistic simulations (fifth column) for Fe$_{50}$Ni$_{50}$ and Py, however for Py$_{60}$Cu$_{40}$ the estimation is not that good. We have to remember that the MFA re-scaling of the exchange parameters did not give a completely satisfactory result for the shape of $m(T)$ in this alloy (see Fig.\ \ref{fig:M-T-Atomistic-MFA}(a)). Thus, since the re-scaled exchange parameter does not work completely well at the low-to-intermediate temperature interval, we further investigate this case (Py$_{60}$Cu$_{40}$) by relating the obtained relation in Eq.\ \eqref{eq:ratio-demag-time-FeNi} for the ratio  $\tau^{\rm{Ni}}/\tau^{\rm{Fe}}$ to the slopes of the curves $m(T)$. 

This can  be easily done by using the linear decrease of magnetization at low temperature, 
$m(T) \approx 1- \kappa  T/\TC$, where $\kappa=W \kB /J_0$ for classical spin models, here $W$ is the Watson integral\cite{dombBOOK2000phase}. Thus, the ratio between the slopes of $m(T)$ for each sublattice is
directly related to the ratio between the exchange values, $\widetilde{J}_0^{\delta}$, 
as follows,
 $\kappa^{\textrm{Fe}}/\kappa^{\textrm{Ni}}=\widetilde{J}_0^{\textrm{Ni}}/
\widetilde{J}_0^{\textrm{Fe}}$.
It is worth noting that  the equilibrium magnetization as a function of temperature can be fitted to the power law  
$m(T)=(1-T/\TC)^{\kappa}$ which in turn gives the low temperature limit 
$m(T) = 1- \kappa  T/\TC$. 
And more importantly, it gives a 
link of the dynamics to the equilibrium thermodynamic properties through the 
ratio 
\begin{equation}
\frac{\tau^{\textrm{Ni}}}{\tau^{\textrm{Fe}}}=
\frac{\lambda^{\rm{Fe}}}{\lambda^{\rm Ni}} 
\frac{\muNI}{\muFE}
\frac{\kappa^{\textrm{Ni}}}
{\kappa^{\textrm{Fe}}}.
 \label{eq:ratio-times-final-expression}
\end{equation}
Next, we fit the numerically evaluated $m(T)$ curves to the power law  $m{\textrm{Fe(Ni)}}(T)=(1-T/\TC)^{\kappa^{\textrm{Fe(Ni)}}}$  for $T < 0.5 \;\TC$.  This allows us to directly estimate the 
ratio between the relaxation times for the three alloys, see Table \ref{table:table2}. 
We can see that the relation in Eq.\ \eqref{eq:ratio-times-final-expression} agrees well  for the 
three alloys even for Py$_{60}$Cu$_{40}$.

For a more general case, for instance at elevated temperatures, where the one-exponential solution is not a good approximation, we have to solve numerically for the coefficients of each 
exponential decay $A^{\epsilon}$ and $B^{\epsilon}$.  
Apart from the exchange interactions and temperature dependence, $A^{\epsilon}$ and $B^{\epsilon}$ also depend on the initial conditions $\delta m^\epsilon (0)=m^{\epsilon}(0)-m_{\rm e}^{\epsilon}$. 

\subsection{Effect on  distinct local damping parameters on the magnetization dynamics}

The intrinsic (atomistic) damping parameters $\lambda^{\epsilon}$ are not be necessarily the same for both sublattices. To investigate the effect of different damping parameters we consider that the magnetic system is initially at equilibrium at room temperature $T=300$ K. Then a heat pulse  $T_{\rm{pulse}}$ is applied for 1 ps.
We define $\tau^{\rm{Fe(Ni)}}$ the time at which the normalized magnetization, $m_{\rm{norm}}(t)= (m(t)-m_{\rm{min}})/({m_{(t = 0)}-m_{\rm{min}}})$ 
is $1/e$. The results for a broad parameter space of $\lambda^{\rm{Fe}}/\lambda^{\rm{Ni}}$ and heat pulse temperature $T_{\rm{pulse}}$ (scaled to $\TC$) are shown in Fig.\\ref{fig:PhaseDiagramRelaxTimesPy}. The line where $\tau^{\rm{Ni}}/\tau^{\rm{Fe}}=1$ lies
at  low pulse temperature  (linear limit in the LLB) $\lambda^{\rm{Fe}}/\lambda^{\rm{Ni}}=1.563$. 
The critical ratio $\left(\frac{\lambda^{\rm{Fe}}}{\lambda^{\rm Ni}}\right)_{\rm{cr}}$ is close to the one which could be predicted  
from  Eq. \eqref{eq:ratio-times-final-expression}  assuming  $\tau^{\rm{Ni}}/\tau^{\rm{Fe}}=1$: 
\begin{equation}
\left(\frac{\lambda^{\rm{Fe}}}{\lambda^{\rm Ni}}\right)_{\rm{cr}} =
\frac{\muFE}{\muNI}
\frac{\kappa^{\textrm{Fe}}}
{\kappa^{\textrm{Ni}}}.
 \label{eq:ratio-times-critical-value}
\end{equation}
  Estimations of this critical ratio at low temperatures  can be found in Table \ref{table:table2}. The ratio is around 2 for all the alloys.

\begin{figure}
\begin{centering}
\includegraphics[scale=0.4, angle = 0]{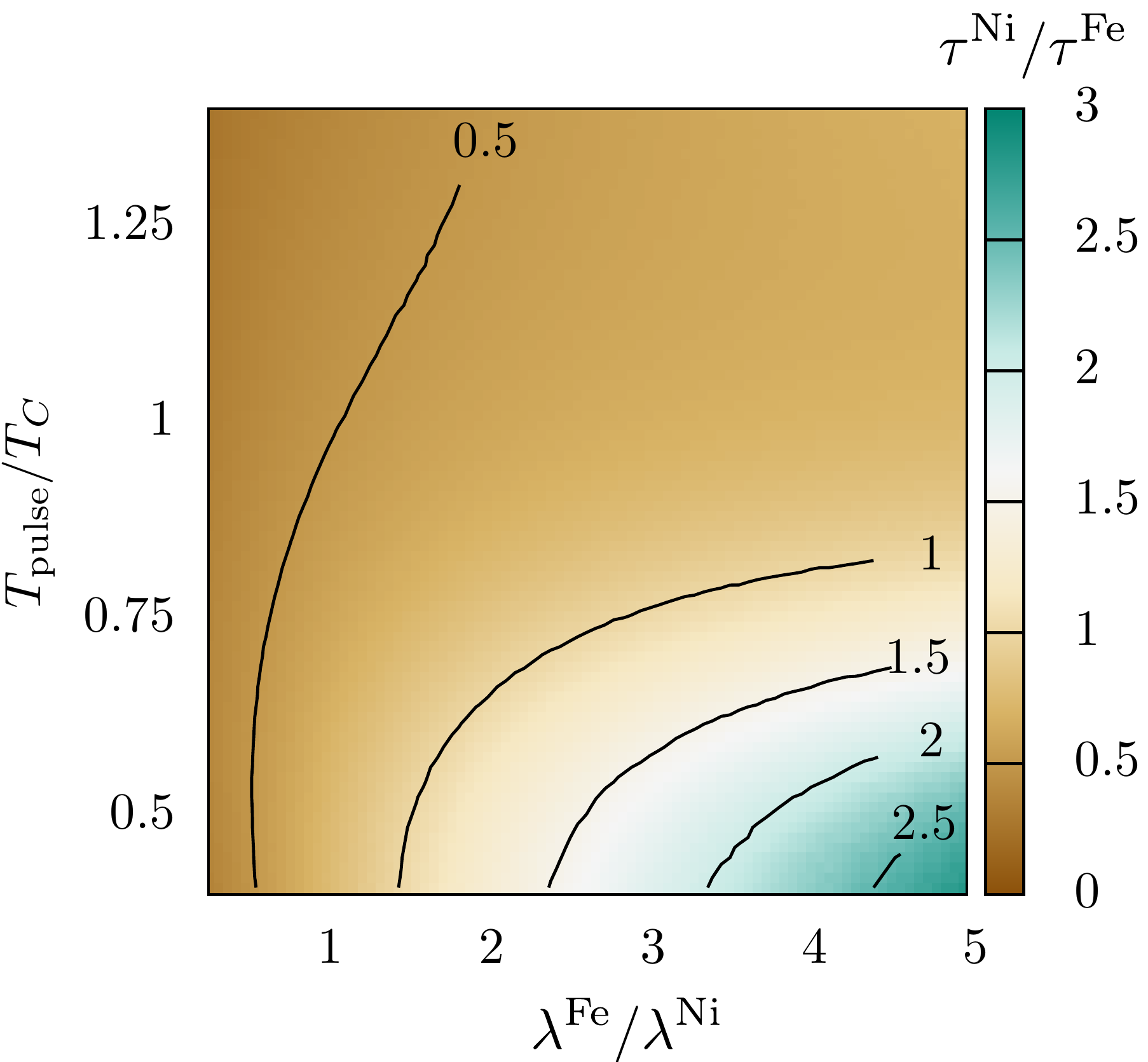}
\end{centering}
\caption{(Color online) Ratio between the relaxation times $\tau$ of the Fe and Ni sublattices in Py after the application of a heat pulse of temperature $T_{\rm{pulse}}$ for a range of values of the ratio of intrinsic damping parameters, $\lambda^{\rm{Fe}}/ \lambda^{\rm{Fe}}$. Black lines represent $\lambda^{\rm{Fe}}/\lambda^{\rm{Ni}}$ values where the ratio between relaxation times  $\tau^{\rm{Ni}}$and 
$\tau^{\rm{Fe}}$ is constant with the value given by the label.} 
\label{fig:PhaseDiagramRelaxTimesPy}
\end{figure}

\textcolor{black}{The results presented in Fig.\ \ref{fig:PhaseDiagramRelaxTimesPy} show a variety of possible situations that can be encountered in experiments on alloys with two magnetic sublattices. They show that in the case of equal coupling to the heat-bath, the Ni sublattice demagnetizes faster than the Fe sublattics in all temperature ranges. The situation may be changed if Fe is  as least twice stronger coupled to the heat-bath than Ni. This conclusion is not inconsistent with the disproportional couplings that were assumed in Ref.\ \onlinecite{SchellekensPRB2013microscopic}. Thus, Fe can demagnetize faster than Ni (as reported in Ref.\ \onlinecite{MathiasPNAS2012probing}) only if Fe is stronger coupled to the heat-bath.}

\section{Discussion and conclusion}

Element-specific magnetization dynamics in multi-sublattice magnets has attracted a lot of attention lately
\cite{chanPRL2009ultrafast,khorsandPRL2013element,MathiasPNAS2012probing}.
The case of GdFeCo ferrimagnetic alloys is paradigmatic since this was the first material where the so-called ultrafast all-optical switching (AOS) of the magnetization has been observed \cite{StanciuPRL2007AOS}.  The element-dependent magnetization dynamics in GdFeCo alloys has meanwhile been thoroughly studied \cite{OstlerNatComm2012ultrafast,MentinkPRL2012ultrafast,BarkerSREP2013two,WienholdtPRB2013orbital,BaryakhtarJETP2013exchange,AtxitiaPRB2013,AtxitiaPRB2014controlling}.
From a fundamental view point, however, it is also important to understand the  element-specific magnetization dynamics  in multi-element ferromagnetic alloys. This is challenging from a modeling perspective and, moreover, contradicting results have been observed in NiFe alloys\cite{MathiasPNAS2012probing,EschenlohrPhD12}.    
 
To treat such alloys  we have developed here a hierarchical multiscale approach for disordered multisublattice ferromagnets. 
The electronic structure \emph{ab initio} calculations of the exchange integrals between atomic spins in FeNi alloys serves as as an accurate foundation to define a classical Heisenberg spin Hamiltonian which in turn has been used to calculate the element-specific magnetization dynamics of atomic spins  through computer simulations based on the stochastic LLG equation. \textcolor{black}{Our simulations predict  consistently a faster demagnetization of the Ni as compared to the Fe. These findings
are  however in contrast to the dynamics measured by Mathias \emph{et al.} \cite{MathiasPNAS2012probing} }

From a modeling perspective, we have linked information obtained from computer simulations of the atomistic Heisenberg Hamiltonian to
large scale continuum theory on the basis of the recently derived  finite temperature LLB model for two sublattice magnets \cite{AtxitiaPRB2012Landau}. 
The LLB model is rather general, it can be applied not only to ferromagnetic alloys, as we have done in the present work, but also to ferrimagnetic alloys. \cite{AtxitiaPRB2014controlling}   
Thanks to analytical expressions coming from the LLB model we have been able to interpret the distinct element-specific dynamics in FeNi alloys in terms of the  strength of the exchange interaction acting on each sublattice. Assuming equal damping parameters for Fe and Ni, the difference is not only coming from the different atomic moments. Analytical expressions derived for the ratio between demagnetization times in Fe and in Ni compare very well to numerical results from computer simulations of the atomistic spin model. To investigate the effect of different intrinsic damping parameters we 
have restrained ourselves to use the LLB approach which is computationally less expensive than the atomistic spin dynamic simulations on a large system of atomic spins.
Our  investigation thus prepares a route to an easier characterization, prediction and hence, control of the thermal magnetic properties of  disordered multi-sublattice magnets, something which will be valuable for technological purposes.

As for the applicability of our multiscale approach to \textit{ferrimagnetic} materials, one would obviously need accurately calculated exchange integrals as a starting point. Computing these  for rare-earth transition metals alloys might not straightforward, as the rare-earth ions contain mostly localized $f$-electrons with a sizable orbital contribution to the atomic moment. However it is expected that for ferrimagnetic alloys, or multilayers with antiparallel alignment,  composed of transition metals this task will be easier. Initial theoretical comparisons of the element-specific demagnetization in GdFeCo were done recently by Atxitia \emph{el al.} \cite{AtxitiaPRB2014controlling} who obtained a  good agreement with experimental observations. However, in this work the exchange integrals as well as the magnetic atomic moments were taken from phenomenological considerations contrary to the present work where all the parameters are obtained from first-principles calculations.

\begin{acknowledgments} 
 This work has been funded through Spanish Ministry of Economy and Competitiveness under the grants MAT2013-47078-C2-2-P, the Swedish Research Council (VR), and by the European Community's Seventh Framework Programme FP7/2007-2013) under grant agreement No.\ 281043, FEMTOSPIN. 
UA gratefully acknowledges support from  EU FP7 Marie Curie Zukunftskolleg Incoming Fellowship Programme, University of Konstanz (grant No.\ 291784). Support from the Swedish Infrastructure for Computing (SNIC) is also acknowledged.
\end{acknowledgments}

\bibliographystyle{apsrev4-1}

\bibliography{Cite}

\end{document}